\documentclass[12pt]{iopart}
\usepackage{graphicx}%

\begin {document}

\title
{ The kinetics of inactivation of spheroidal microbial cells by
pulsed electric fields}
%\author {Nikolai I.  Lebovka$^1,2$ and
%Eugene Vorobiev$^2$ } \affiliation {
%$^1$F. D. Ovcharenko Biocolloid Chemistry Institute, 42 Vernadsky Av., Kyiv, Ukraine\\
%$^2$Departement de Genie Chimique, Universite de Technologie de
%Compiegne, Centre de Recherche de Royallieu, B.P. 20529-60205
%Compiegne Cedex, France}

\author{Nikolai I Lebovka$^1$, Eugene Vorobiev$^2$, }

\address{$^1$ Institute of Biocolloidal Chemistry named after F. D.
Ovcharenko, NAS of Ukraine, 42, blvr. Vernadskogo, Kiev, 03142,
Ukraine}

\address{$^2$ Departement de Genie Chimique, Universite de Technologie de
Compiegne, Centre de Recherche de Royallieu, B.P. 20529-60205
Compiegne Cedex, France}

\date{\today}%

\begin{abstract}
The nature of non-exponential kinetics in microbial cells
inactivation by pulsed electric fields (PEF) is discussed. It was
demonstrated that possible mechanism of non-exponential kinetics
can be related to orientational disorder in suspension of
microbial cells of anisotropic form. A numerical studies of
spheroidal cell suspensions was carried out. The most pronounced
deviations from the exponential kinetics were observed for
disordered suspensions of prolate spheroids at small electric
field strength $E$ or at large aspect ratio $a$. For partially
oriented suspensions,  efficiency of inactivation enhances with
increasing of order parameter and field strength. A possibility of
the PEF-induced orientational ordering in microbial suspensions is
discussed.
\end{abstract}
% line{\pacs {87.00.00; 87.16.Dg; 87.17.Aa; 87.50.Rr; 89.75.Da}}
%\keywords{Electric fields; Kinetics of microbial inactivation; Spheroidal cells; Transmembrane voltage; Orientational ordering}
\pacs{87.00.00; 87.16.Dg; 87.17.Aa; 87.50.Rr; 89.75.Da}
%\submitto{\}

\maketitle

%87.00.00 Biological and medical physics
%87.16.Dg Membranes, bilayers, and vesicles
%87.17.-d Cellular structure and processes
%87.17.Aa Theory and modeling; computer simulation
%83.80.Hj Suspensions, dispersions, pastes, slurries, colloids
%87.50.Rr Electric fields

\section{Introduction}

Application of high intensity pulsed electric field (PEF) as a
nonthermal preservation method have been a topic of growing
interest over the past decade \cite{Jeyamkondan1998,Hoover1997}.
PEF processing is a promising non-thermal alternative technology
and many investigators have shown it effectiveness for killing
bacteria in liquid aqueous media
\cite{Barbosa-Canovas1998,Barsotti1998,Vega-Mercado1998,Wouters1997}.
The application of high electric field microsecond pulses
(typically 10-50 kV/cm) allows inactivation of microorganisms and
enzymes
\cite{Sizer1999,Lado2002,Teissie2002,Ross2003,Devlieghere2004}.
However, the mechanism of inactivation are not yet fully
understood. The important problem is to elucidate how kinetics and
degree of inactivation depend on the type and geometrical
parameters of microorganisms, properties of liquid media and their
flow regimes, the temperature and treatment protocol (electric
field strength $E$, wave forms, pulse duration $t_i$, total time
of treatment) \cite{Qin1996,Eynard2003}.

The microbicidal effect of PEF treatment is related to selective
damage of the biological membrane in microorganism. In an external
electric field $E$, a transmembrane potential $u_m$ is induced on
membrane. When transmembrane potential exceeds some threshold
value (typically about 0.2-1.0 V), an electric field cause a
temporary loss of the semipermeability of cell membranes or their
electroporation \cite{Weaver1996, Chen2006}. The sufficiently
strong electric field and long time of PEF treatment leads to
complete membrane damage and cell death \cite{Pliquett2007}.

The surviving fraction $S(t,E)$, defined as the ratio of the
number of undamaged microbial cells to the total number of
microbial cells, decreases with PEF treatment time $t$. Different
empirical models, such as Fermi, Hulsheger, log-log and
log-logistic, were proposed for description of inactivation
kinetics \cite{Barbosa-Canovas1998,Hulsheger1983,Peleg1996,
Alvarez2000}. Although these models are very popular, they have no
theoretical justification. The Weibull distribution was
successfully applied in a number of works for fitting experimental
PEF inactivation data for a surviving fraction $S(t,E)$
\cite{Smelt2002,
Alvarez2003a,Alvarez2003b,Alvarez2003c,Rodrigo2003}
\begin{equation}
S(t,E) = \exp(-(t/\tau(E))^{n(E)}). \label{e1}
\end{equation}

Here, $\tau ( {E} )$ and $n ( {E}  )$ are the time and shape
parameters accounting for the effective inactivation time and
survival curve concavity, respectively \cite{vanBoekel2002}. But
the Weibull model is also empirical and physical meaning of the
obtained parameters $\tau(E)$ and $n(E)$ was not elucidated yet.
Lebovka and Vorobiev \cite{Lebovka2004} proposed a theoretical
model for description of the surviving curves of
spherically-shaped bacteria with the cell size distribution. It
was shown that the Weibull model can be successfully applied for
fitting of the surviving curves during PEF treatment.

Microbial cells display a variety of shapes and dimensions
\cite{Bergey1986,Freitas2001} depending on the culture condition
and age. Size of cells varies between 0.1 $\mu$m and 10 $\mu$m.
Generally, the following different shapes can occur:
near-spherical, ellipsoidal or ovoid (cocci), cylindroidal
(bacilli), and spiral or comma-like (spirilli). For example, cells
of \textit{Escherichia coli} and \textit{Salmonella typhi} are
rod-like and have 0.4-0.6 $\mu $m in diameter and 2 - 4 $\mu $m in
length, cells of\textit{ Leptospira spp.}  are very long rod with
0.1 $\mu $m in diameter and 20 $\mu $m in length, cells\textit{}
of\textit{ Staphylococcus} \textit{spp} are spherical cells with
diameter of 0.5-1.5 $\mu $m, \textit{Sacchoromyces cerevisiae}
have ellipsoid cells with the principal dimensions of 2-8 $\mu $m
and 3-15 $\mu $m, respectively, cells of \textit{Klebsiella
pneumoniae} are ovoid with a mean dimension 0.4 $\mu $m and
\textit{Vibrio cholerae} have comma-like cells with the principal
dimensions of 0.5 $\mu $m and 1.5-3 $\mu $m, respectively
\cite{Bergey1986,Freitas2001}.

In general case, the surviving kinetics may be rather complex. An
electroporation could be influenced by the aggregation of cells,
their arrangement, local cell density, local solute concentration,
and distribution of local electric field
\cite{Pavlin2002,Canatella2004, Pucihar2007, Pavlin2007}. The
killing probability of non-spherical cells depends substantially
on spatial orientation and changes from cell to cell
\cite{Teissie2002,Heinz2001,Valic2003} and can be related with
cell diameters, spatial and orientational distributions.

 In this work, a theoretical model allowing to describe the survivor curves
for disordered or partially oriented non-spherical bacteria is
formulated. The model predict how the cell orientation influences
the lifetime of a spheroidal microbial cell exposed by PEF. The
numerical simulations of surviving kinetics of disordered and
partially ordered suspensions of microbial cells were done.

\section{Description of the model}

\subsection{Transmembrane potential}

In general case, electroporation consists of different stages
including the charging of the membrane, creation of pores and
evolution of pore radii \cite{Krassowska2007}. For a single
spherical cell under the steady state conditions, the
transmembrane potential depends on the angle $\varphi$ between the
external field $E$ direction and the radius-vector \textit{r} on
the membrane surface \cite{Schwan1957}:
\begin{equation}
u_{m} = 1.5fRE\cos\varphi. \label{e2}
\end{equation}

Here, $R$ is the cell diameter, and $f$ is a parameter depending
on electrophysical and dimensional properties of the membrane,
cell and surrounding media. In dilute suspensions of cells, the
value of $f$ is close to 1.

The value of $u_{m} $ is proportional to the cell radius $R$. The
highest drop of potential occurs at the cell poles and decreases
to zero at $\varphi = \pm \pi /2$. So, the larger microbial cells
get killed before smaller ones and the damage probability is
maximal at the cell poles.

If a cell is non-spherical, the transmembrane potential $u_m$
becomes more complex function of the cell size and geometry,
direction of external field and position on the membrane surface.
The transmembrane potential $u_m$ of an arbitrary oriented
ellipsoidal cell at some point on the membrane surface $r(x,y,z)$
may be calculated from the following generalized Schwan equation
\cite{Fricke1953}:
\begin{equation}
u_m = \sum\limits_{i= x,y,z} {r_i E_i / ( {1 - L_{i}}
 )}. \label{e3}
\end{equation}

Here, $L_i$ are the depolarising factors defined by the cell radii
$R_x$, $R_y$ and $R_z$ \cite{Landau1984}. This approximation works
for a membrane with negligibly small conductance and its
application was discussed extensively in literature
\cite{Valic2003,Bernhardt1973, Zimmermann1974,Hart1982,Kotnik2000,
Gimsa2001}.

The depolarizing factor of a prolate spheroid ($R_z>R_x=R_y$)
along the symmetry axis $z$ is \cite{Landau1984}
\begin{equation}
L_{z} = \frac{{1 - e^{2}}}{{2e^{3}}} ( {\ln\frac{{1 + e}}{{1 - e}}
- 2e}  ), \quad e = \sqrt {1 - a^{ - 2}} , \label{e4a}
\end{equation}
and for an oblate spheroid ($R_z<R_x=R_y$) it makes
\begin{equation}
L_{z} = {{{1} \over {e^{3}}}} ( {e - \sqrt {1 - e^{2}\arcsin e}}
 ), \quad e = \sqrt {1 - a^{2}} , \label{4b}
\end{equation}
where $a=R_z/R_x$ is an aspect ratio (major $/$ minor axis). The
depolarizing factors in $x$ and $y$ directions are defined as
\begin{equation}
L_x = L_y =  ( 1 - L_z)/2.\label{e5}
\end{equation}

Here, $L_x=L_y=L_z=1/3$ for a spherical cell, $L_x=L_y \approx
0.5$, $L_z \approx 0$ for a long cylinder and $L_x=L_y \approx 0$,
$L_z \approx 1$ for a thin disk.

\begin{figure}[h]
\begin{center}
\includegraphics[width=12.0cm,clip=true]{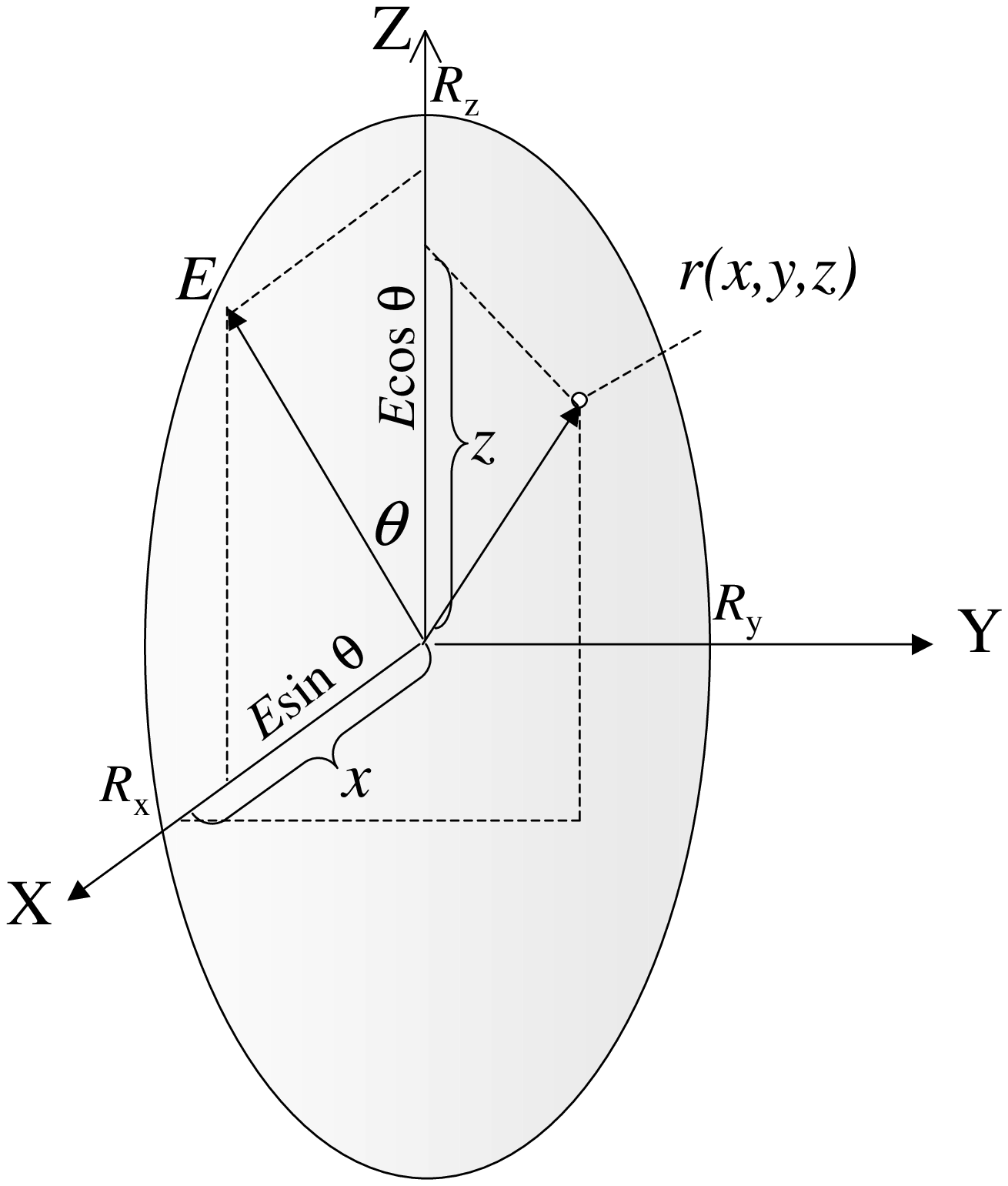}
\end{center}
\caption{Schematic representation of the problem under
consideration. $X,Y,Z$ are the local Cartesian coordinates related
to the oblate spheroidal microbial cells in the external field $E$
(it is supposed here that it is lies in the $XOZ$ plane). Here,
$\theta $ is an angle between the electric field direction $E$ and
symmetry axis of spheroid $Z$, $r(x,y,z)$ is the radius of a
membrane surface point, where the transmembrane potential is
calculated, $R_{x}=R_{y}, R_{z}$ are the cell radii. } \label{f01}
\end{figure}

In general case, the transmembrane potential can be calculated
from \eref{e3}, but a simpler form of this equation may be
considered due to the symmetry of spheroid \cite{Valic2003}. When
the electric field vector lies in a $X0Z$ plane (see \fref{f01}),
\eref{e3} can be rewritten as
\begin{equation}
u_{m} = x E\sin\theta / ( {1 - L_{x}}   ) + z E\cos\theta / ( {1 -
L_{z}}  ), \label{e6}
\end{equation}
where $\theta$ is an angle between the external field and symmetry
axis of spheroid, and $x$ and $z$ are coordinates of a point at
the spheroid surface. The values of $x$ and $z$ in a spheroidal
system are defined as \cite{Korn2000}
\begin{equation}
x = R_{x} \sqrt {1 - \eta ^{2}} \cos\phi, z = R_{z} \eta,
\label{e7}
\end{equation}
where $-1 \le \eta  \le 1$, $0 \le \phi  \le 2\pi $ are the
spheroidal coordinates.

Finally, introducing \eref{e7} into \eref{e6}, we obtain

\begin{equation}
u_m = ERa^{ - 1/3} ( \frac{\sin\theta\cos\phi\sqrt{1 - \eta ^2}}{1
- L_x} + \frac{\cos\theta a\eta}{{1 - L_{z}}}) , \label{e8}
\end{equation}
where $R$ is a radius of sphere, that has the same volume $V$ as
spheroid ($V =4\pi R^3/3=4\pi R_z R_x^2/3$).

\subsection{Lifetime of a microbial cell exposed by PEF}

The lifetime of a membrane in some point at the spheroid surface
can be estimated on the basis of the transient aqueous pore model
 \cite{Weaver1996}:
\begin{equation}
\tau(u_m(\theta ,\eta ,\phi)) = \tau _{\infty} \exp\frac{{\pi
\omega ^{2}/kT\gamma} }{{1 +  ( {u_{m}  ( {\theta ,\eta ,\phi}
)/u_{o}}   )^{2}}}. \label{e9}
\end{equation}

Here, $\tau _{\infty}  $ is the parameter ($\tau \to \tau
_{\infty} $ in the limit of very high electric fields), $\omega $
and $\gamma $ are the line and surface tensions of a membrane,
respectively, $k = 1.381^{.}10^{-23}$ J/K is the Boltzmann
constant, $T$ is the absolute temperature, $u_o = \sqrt {2\gamma
/( {C_{m}  ( {\varepsilon _{w} /\varepsilon _{m} - 1} )} )}$ is
the voltage parameter (the dimension of $u_{o}$ is Volts), $C_{m}
$ is the specific capacitance of a membrane, $\varepsilon _{w},
\varepsilon _{m} $ are the relative dielectric permittivities of
the aqueous phase and of the membrane, respectively.

The lifetime of a spheroidal cell $\tau _{c}$ depends on the angle
$\theta $ between electric field direction $E$ and the symmetry
axis $Z$ of a spheroid. It can be estimated by a averaging of
$\tau ^{-1}(u_m)$ on the spheroid surface:

\begin{equation}
\tau _c^{-1}( \theta) = R_{x}^{2} A^{-1}\int\limits_{-1}^{1}
{\int\limits_{0}^{2\pi} \frac{\sqrt {a^{2} ( {1 - \eta ^{2}}  ) +
\eta ^{2}}}{{\tau ( {u_{m}  ( {\theta ,\eta ,\phi}   )}  )}  }
}d\eta d\phi , \label{e10}
\end{equation}
where
\begin{equation}
A = 2\pi R_{x}^{2} \int\limits_{ - 1}^{1} {\sqrt {a^{2} ( {1 -
\eta ^{2}}  ) + \eta ^{2}} d\eta} \label{e11}
\end{equation}
is the surface area of a spheroid \cite{Korn2000}.

For a prolate spheroid, the surface area is
\begin{equation}
A = 2\pi R_x^2 ( 1 +\frac{a\arcsin \sqrt{1-a^{-2}}}{\sqrt
{1-a^{-2}}}), \label{e12a}
\end{equation}
and for a oblate spheroid it makes
\begin{equation}
A = 2\pi R_x^2 ( 1 +\frac{a\arcsin h\sqrt{a^{-2}-1}}{\sqrt
{a^{-2}- 1}}). \label{e12b}
\end{equation}

\subsection{Surviving probability during a PEF treatment}

A surviving probability of a single spheroid with the angle
$\theta $ of its principal axis relative to the external electric
field $E$ is defined as
\begin{equation}
S ( t,\theta) = \exp (- t/\tau_c(\theta)). \label{e13}
\end{equation}

Then, a surviving probability $S(t)$ of the whole suspension with
spheroids of different spatial orientation can be calculated as:
\begin{equation}
S(t) = \int\limits_{-1}^1 f(\theta)\exp(-t/\tau_c
(\theta))d\cos\theta, \label{e14}
\end{equation}
where $f(\theta)$ is an angular orientational distribution
function of spheroids.

For randomly oriented spheroids $f(\theta )=1/2$. For partially
oriented spheroids it is useful to introduce an order parameter
$Q$ defined as \cite{Levy2003}:
\begin{equation}
Q = \frac{1}{2} \int\limits_{-1}^1f(\theta)(3\cos^2\theta-1
)d\cos\theta. \label{e15}
\end{equation}

For perfectly oriented suspension, when all spheroids are
completely aligned, $Q=1$ and for randomly oriented suspension the
order parameter is zero, $Q=0$.

Disordered suspensions of anisotropic cells may be oriented by the
external electric or magnetic fields \cite{OKonski1976,
Stoylov1991,Schafer1998, Rudakova1999}, or by the fluid flow
\cite{Matsumoto1999, Vernhes2002, Teissie2002}. In the external
electric field $E$ the angular orientation distribution function
$f(\theta )$ can be estimated as \cite{Stoylov1977, Levy2003}
\begin{equation}
f ( {\theta}   ) = \frac{{\exp ( {U^{\ast} \cos^{2}\theta}
 )d\cos\theta} }{{\int\limits_{ - 1}^{1} {\exp ( {U^{\ast}
\cos^{2}\theta}   )d\cos\theta} } }, \label{e16}
\end{equation}
where $U^*$  is a dimensionlees electrostatic energy of spheroid
in the external field $E$,
\begin{equation}
U^{\ast}  = \beta E_{}^{2} / ( {2kT}  ). \label{e17}
\end{equation}

Here, $\beta $ is the electrical polarizability anisotropy of a
particle that depends on electrophysical properties of the
particle and the outer solution.

An order parameter $Q$ can be calculated by substitution of
\eref{e16} into \eref{e15} \cite{Konski1959}
\begin{equation}
Q = \frac{{3}}{{4\sqrt {U^{\ast} }} } ( {\exp ( {U^{\ast} }
 )/\int\limits_{0}^{\sqrt {U^{\ast} }}  {\exp ( {t^{2}}
 )dt} - 1/\sqrt {U^{\ast} }}   ) - 1/2. \label{e18}
\end{equation}

Proceeding from \eref{e14}-\eref{e18}, surviving kinetics versus
order parameter $Q$ can be calculated for partially ordered
microbial cells.

\subsection{Details of numerical calculations}

For $\tau_c(\theta)$ evaluation using \eref{e8}-\eref{e11}, the
double integration was done using Simpson's quadrature rule. The
accuracy of numerical integration was better than $10^{-6}$. The
voltage scale parameter in \eref{e9} was estimated as $u_{o}
\approx 0.17$V from data obtained by Lebedeva \cite{Lebedeva1987}
for the general lipid membranes ($\omega \approx 1.69\ast 10^{ -
11}$N, $\gamma \approx 2\ast 10^{ - 3}$ N/m, $\varepsilon _{w}
\approx 80$, $\varepsilon _{m} \approx 2$, $C_{m} \approx 3.5\ast
10^{ - 3}$F/m$^{2}$ at $T = 298$K). The time scale parameter was
put as $\tau _{\infty} \approx 3.7\ast 10^{ - 7}$s
\cite{Lebedeva1987}. It is useful to use in calculations a
dimensionless reduced field intensity defined as $E^{\ast } =
E/E_{o} $, where $E_{o} = 2u_{o} /3R$, and $R$ is an equivolume
radius of a spheroid. Note that at $2R \approx 1\mu$m, and $u_{o}
\approx 0.17$V, $E_{o} \approx 2.27$kV/cm. All these parameters
were used calculation for estimation purposes.

The calculated dependencies of $\tau _{c}$($\theta $) were used
for numerical calculation of the surviving kinetics from
\eref{e14} at different values of order parameters $Q$
(\eref{e18}).

\section{Results and discussion}

\subsection{Lifetime of a spheroidal microbial cell}

\begin{figure}[h]
\begin{center}
\includegraphics[width=12.0cm,clip=true]{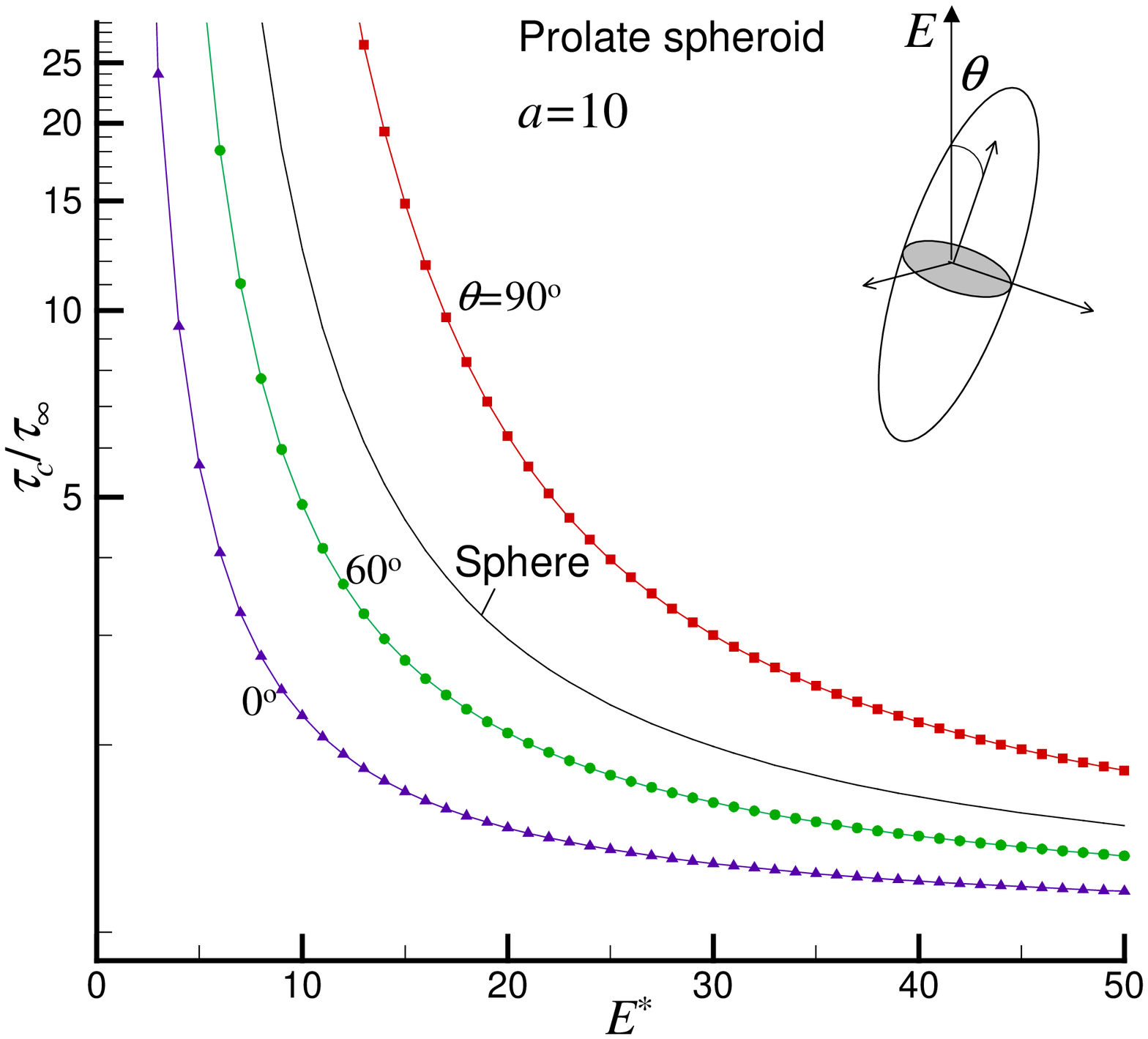}
\end{center}
\caption{Relative lifetime $\tau _{c}$/$\tau_\infty$ of a prolate
(aspect ratio $a=10$) cells versus reduced field intensity $E^*$
at different angles $\theta $ between electric field direction and
symmetry axis of spheroids. The solid lines show data for a
spherical cell with a radius equivalent to that of an equivolume
spheroid.} \label{f02a}
\end{figure}

\begin{figure}[h]
\begin{center}
\includegraphics[width=12.0cm,clip=true]{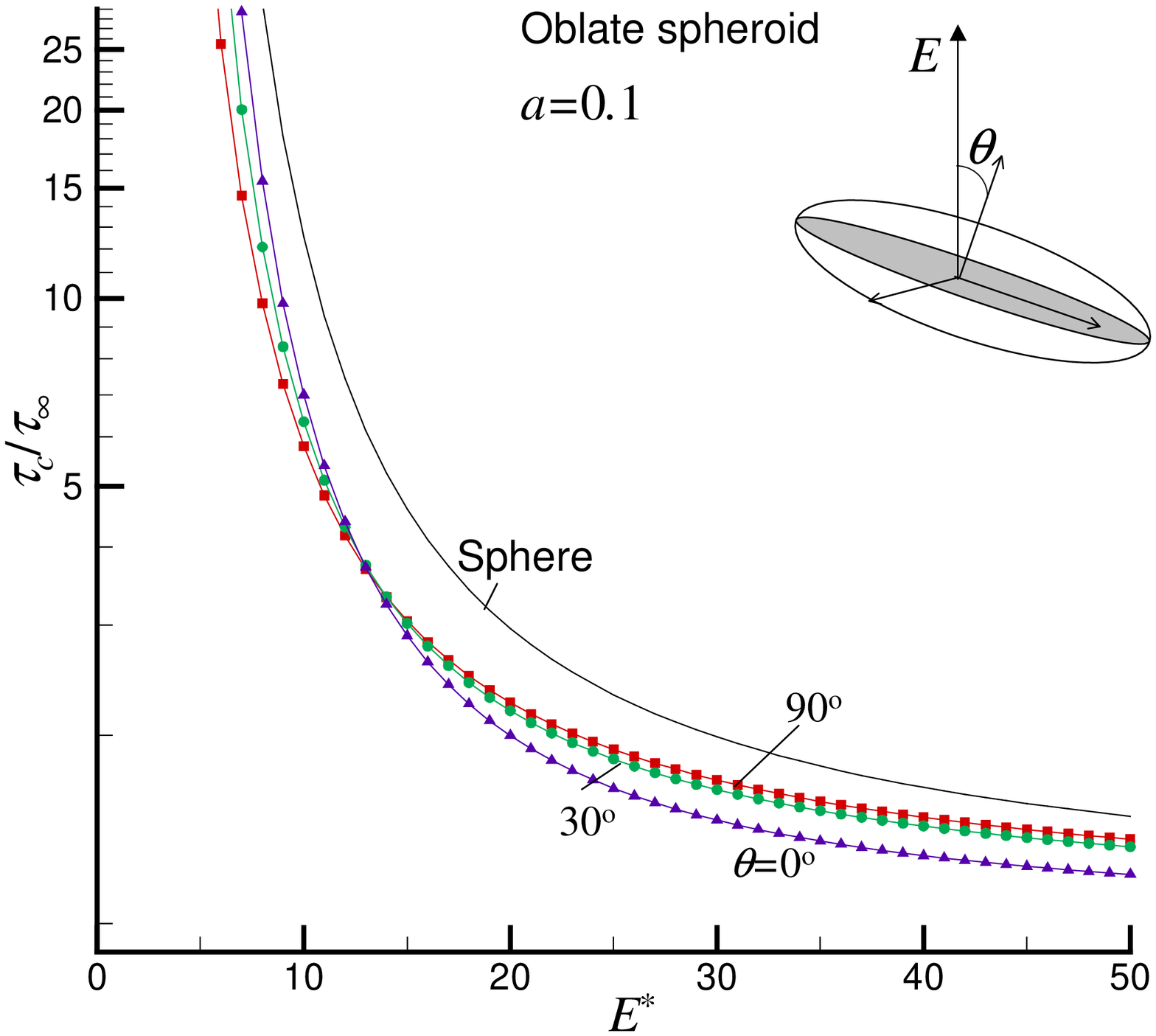}
\end{center}
\caption{Relative lifetime $\tau _{c}$/$\tau_\infty $ of a
 oblate (aspect ratio $a=0.10$) cells
versus reduced field intensity $E^*$ at different angles $\theta $
between electric field direction and symmetry axis of spheroids.
The solid lines show data for a spherical cell with a radius
equivalent to that of an equivolume spheroid.} \label{f02b}
\end{figure}

\Fref{f02a} and \fref{f02b} present some examples of the
calculated relative lifetime  $\tau _c/\tau_\infty$ versus reduced
field intensity $E^*$ for a prolate (\ref{f02a}) and oblate
(\ref{f02b}) spheroids at different values of angle $\theta $. A
prolate spheroid in external electric field was more stable at
$\theta =90^o$ and less stable at $\theta=0^o$ than a spherical
cell of the same volume, but an oblate spheroid was always less
stable electrically than a spherical cell of the same volume.

\begin{figure}[h]
\begin{center}
\includegraphics[width=12.0cm,clip=true]{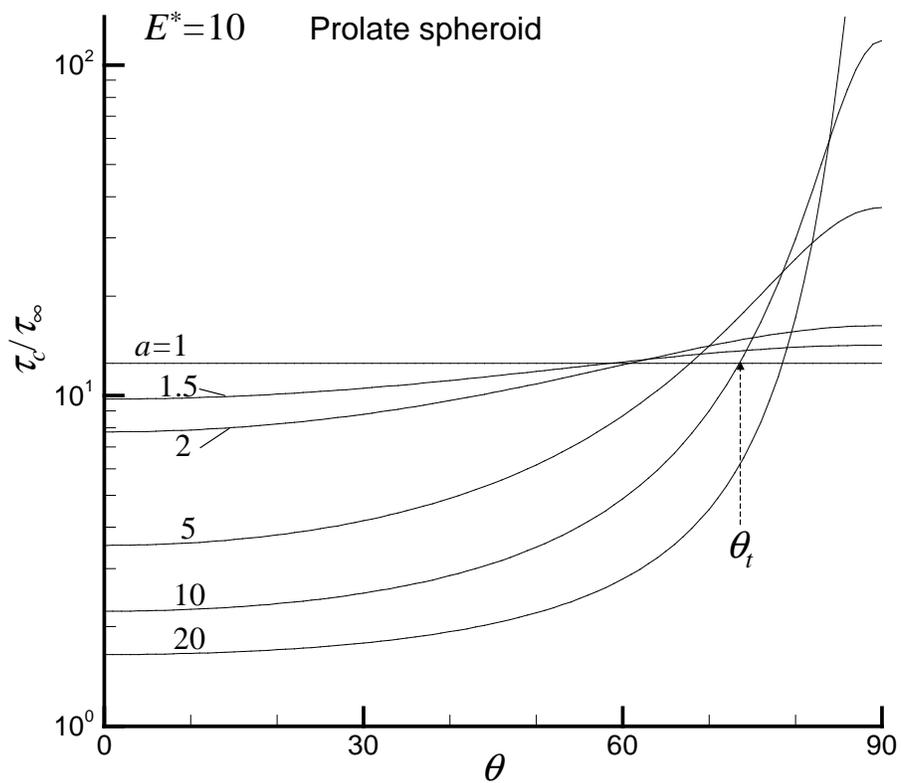}
\end{center}
\caption{Relative lifetime $\tau _{c}$/$\tau_\infty $ of a prolate
spheroidal cell versus angle $\theta $ for different aspect ratio
$a$. The calculation were done at the given value of reduced
electric field intensity $E^*=10$ that corresponds to $E \approx
22.7$ kV/cm for an equivolume radius of spheroid $2R \approx 1\mu
$m, and voltage parameter $u_{o} \approx 0.17$V
\cite{Lebedeva1987}. Arrows show threshold angles $\theta _{t}$.}
\label{f03a}
\end{figure}

\begin{figure}[h]
\begin{center}
\includegraphics[width=12.0cm,clip=true]{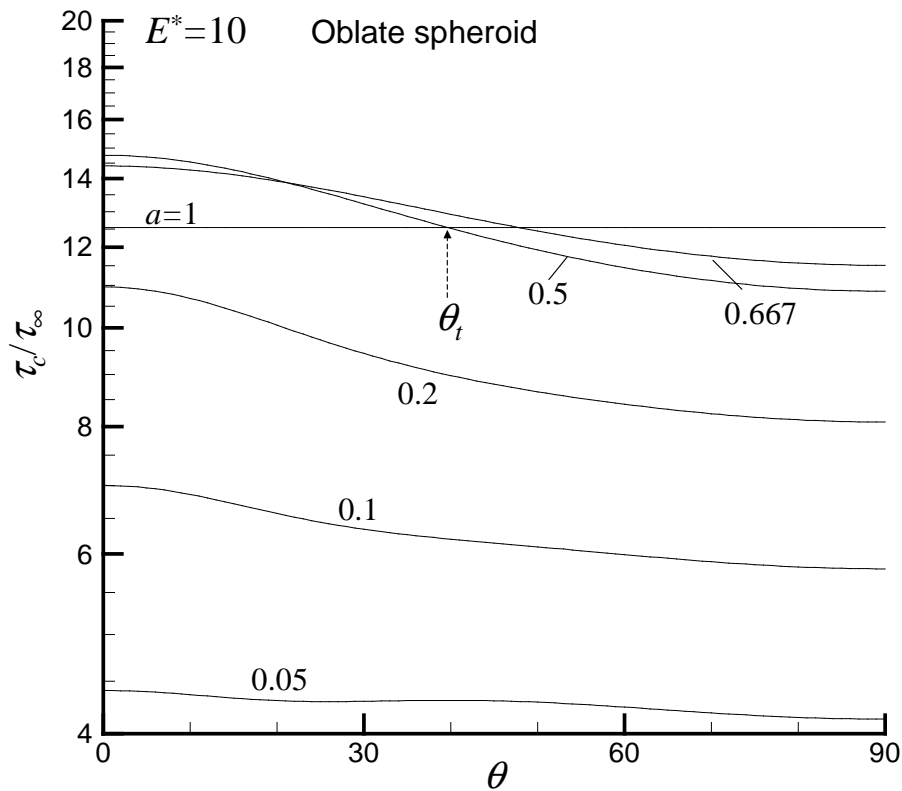}
\end{center}
\caption{Relative lifetime $\tau _{c}$/$\tau_\infty $ of a oblate
spheroidal cell versus angle $\theta $ for different aspect ratio
$a$. The calculation were done at the same condition as for data
in \fref{f03a}. Arrows show threshold angles $\theta _{t}$.}
\label{f03b}
\end{figure}

The relative lifetime $\tau _{c}$/$\tau_\infty$ versus angle
$\theta $ for different aspect ratio $a$ for a prolate  and oblate
 spheroids at \textit{E}*=10 are presented in \fref{f03a} and \fref{f03b}.

For a prolate spheroid, the value of $\tau _{c}$/$\tau_\infty$
considerably increases with angle $\theta $ increase and it was a
minimum for a cell aligned along the applied field $E$
(\fref{f03a}). This result is in accordance with experimental
observations of \cite{Valic2003}, who reported minimum
electropermeabilization for the cells aligned along to the
electric field direction.

For an oblate spheroid, the value of $\tau_c/\tau_\infty$ was
smaller than for a spherical cell of the same volume and the value
of $\tau_c/\tau_\infty$ decreases with angle $\theta$ increase
(\fref{f03b}). It was maximal for $\theta=0$, but
$\tau_c/\tau_\infty$ dependence versus angle $\theta $ was not so
distinct as for a prolate spheroid.

There exist some threshold angle $\theta=\theta_t$, at which the
curve $\tau_c/\tau_\infty(E^*)$ for a prolate spheroid is very
close to that for a spherical cell of same volume. The cell
permeabilization was suppressed at $\theta>\theta_t$ and at was
enhanced $\theta<\theta_t$ as compared with a spherical cell of
the same volume. The higher was the aspect ratio $a$, the larger
was the threshold angle $\theta_{t}$. For an oblate spheroid, the
threshold angle $\theta=\theta_t$ was observed only at high aspect
ratio $a>0.3$.

\subsection{Survivor kinetics of a disordered suspension of cells}

\begin{figure}[h]
\begin{center}
\includegraphics[width=12.0cm,clip=true]{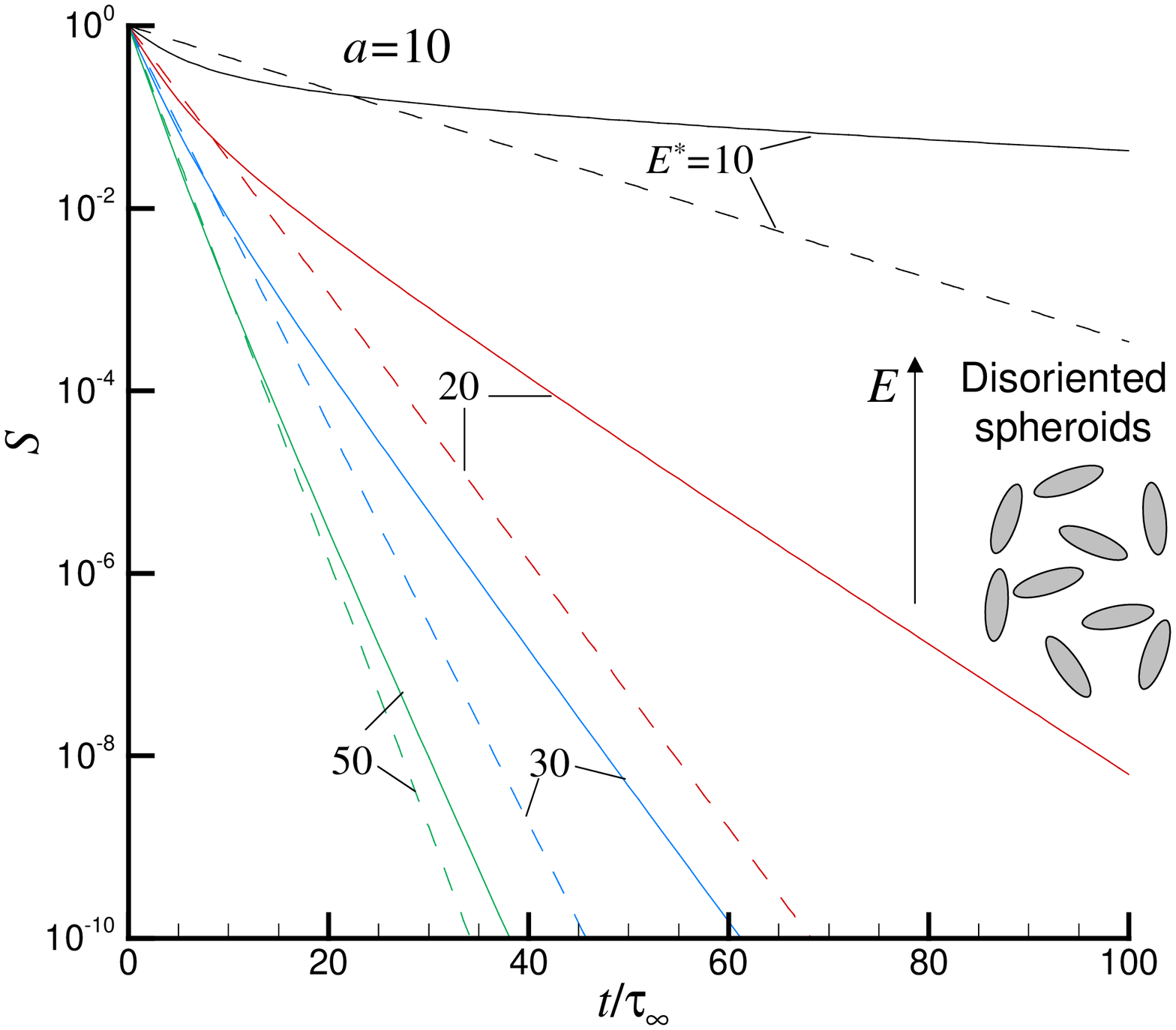}
\end{center}
\caption{Survivor curves $S ( {t/\tau _{\infty} }   )$ for
orientationally disordered spheroids at given $a=10$ and different
$E^{*}$. The dashed lines show data for spherical cell equivolume
with spheroids.} \label{f04a}
\end{figure}

\begin{figure}[h]
\begin{center}
\includegraphics[width=12.0cm,clip=true]{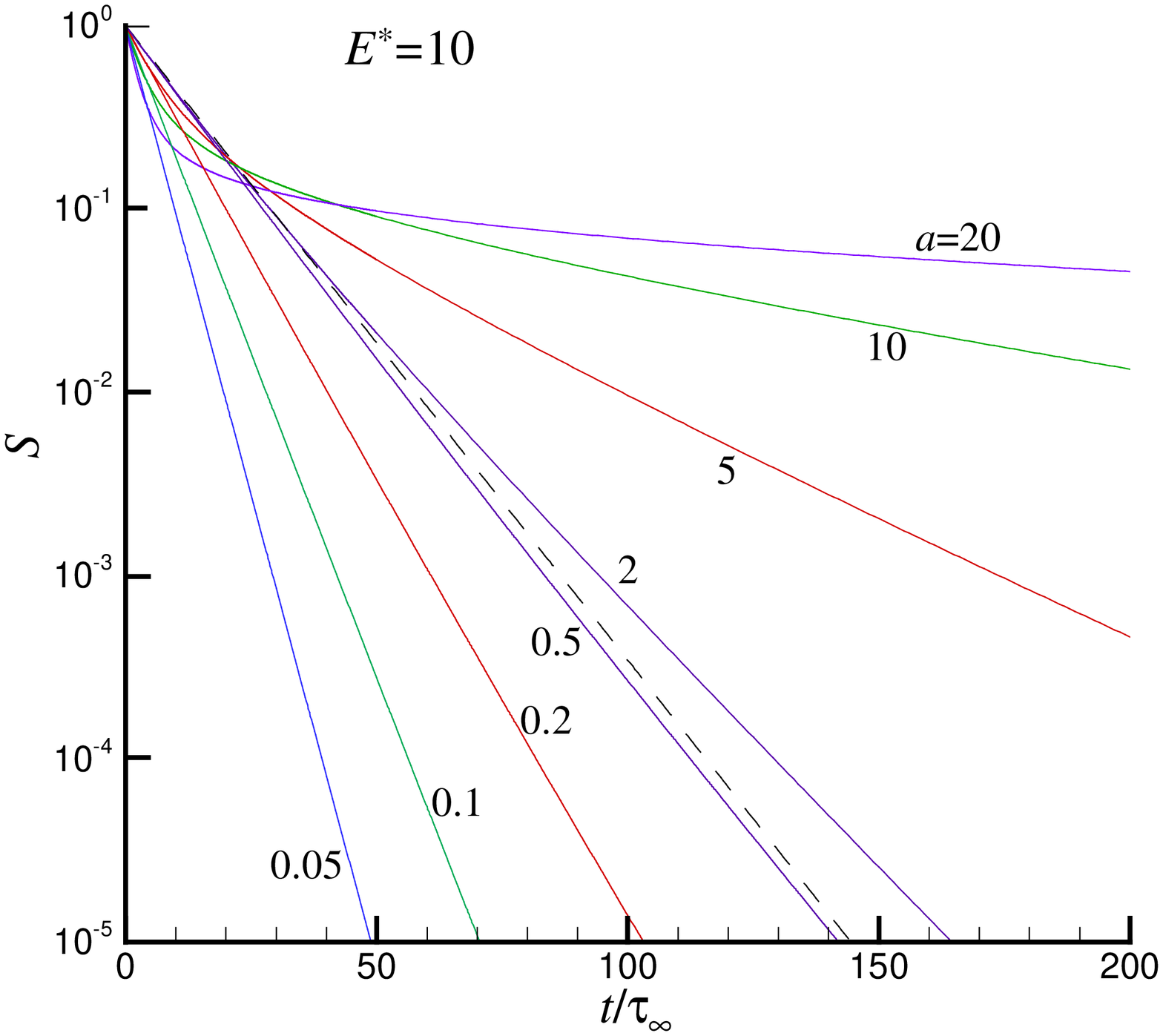}
\end{center}
\caption{Survivor curves $S ( {t/\tau _{\infty} }   )$ for
orientationally disordered spheroids at given $E^{*}$=10 and
different $a$ . The dashed lines show data for spherical cell
equivolume with spheroids.} \label{f04b}
\end{figure}

Due to the Brownian motion a random orientational distribution for
suspensions of microbial cells is typical when field-induced
ordering effects are absent. \Fref{f04a} and \fref{f04b} show the
calculated survivor curves $S(t)$ of disordered suspensions
($f(\theta )=1/2$) in \eref{e14}). The ideal first order kinetics
law \eref{e13} was observed for suspensions of identical spherical
cells (dashed lines in \fref{f04a} and \fref{f04b}). The
deviations from first order kinetics for prolate spheroidal cells
became more pronounced with decreasing of the electric field
intensity $E^*$ (\fref{f04a}) or increasing of the aspect ratio
$a$ (\fref{f04b}). The $\tau_{c}/\tau_\infty$ versus $\theta $
dependence was not so pronounced for oblate cells as for prolate
cells, and no noticeable deviations from the first order kinetics
were observed.

The kinetics $S(t)$ demonstrates that surviving probability in
disordered suspension was higher for prolate cells and was lower
for oblate cells as compared with the surviving probability for
spherical cells of equivalent volume (\fref{f04b}).

The calculated survivor curves $S (t)$ for the prolate cells may
be fitted with empirical Weibull function \eref{e1}. This model
always gives only upward concavity, i.e. $n<1$, for
orientationally disordered suspensions of prolate cells. But the
numerically estimated shape $n$ and relative time $\tau
$/$\tau_\infty $ parameters were rather sensitive to the upper
cutting boundary $t_{max}/\tau_\infty$. This fact reflects
existence of an intrinsic inconsistency between an unknown
survival function and Weibull function  \cite{Lebovka2004}.

\subsection{Survivor kinetics of a partially oriented suspension of cells}

\begin{figure}[h]
\begin{center}
\includegraphics[width=12.0cm,clip=true]{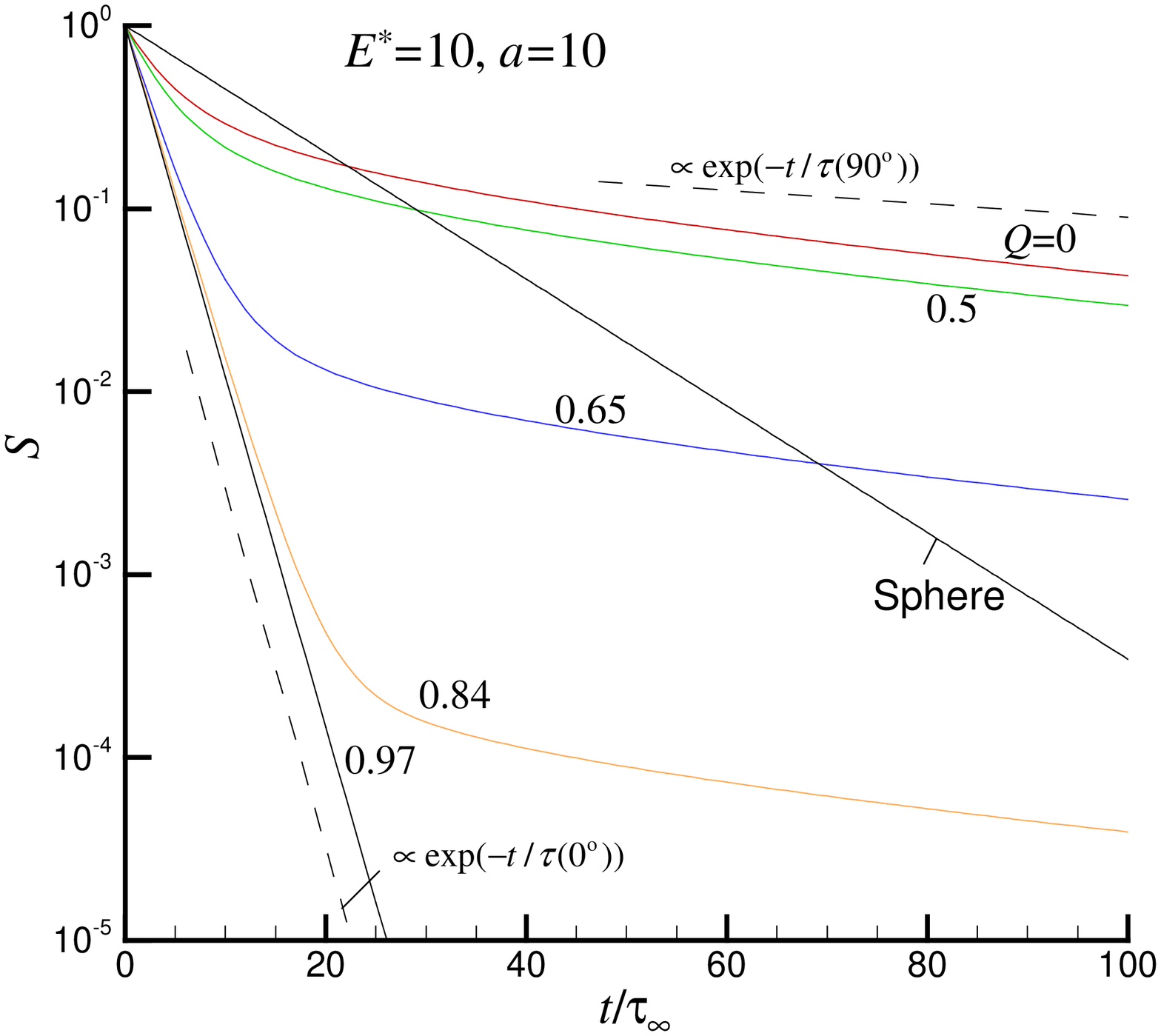}
\end{center}
\caption{Survivor curves $S ( {t/\tau _{\infty} }   )$ for
partially ordered suspensions of prolated spheroids at $a=10$,
$E^{*}=10$ and different $Q$ values. The dashed lines show the
slopes that correspond to orientation with $\theta $=0 and $\theta
$ =90$^{o}$.} \label{f05a}
\end{figure}

\begin{figure}[h]
\begin{center}
\includegraphics[width=12.0cm,clip=true]{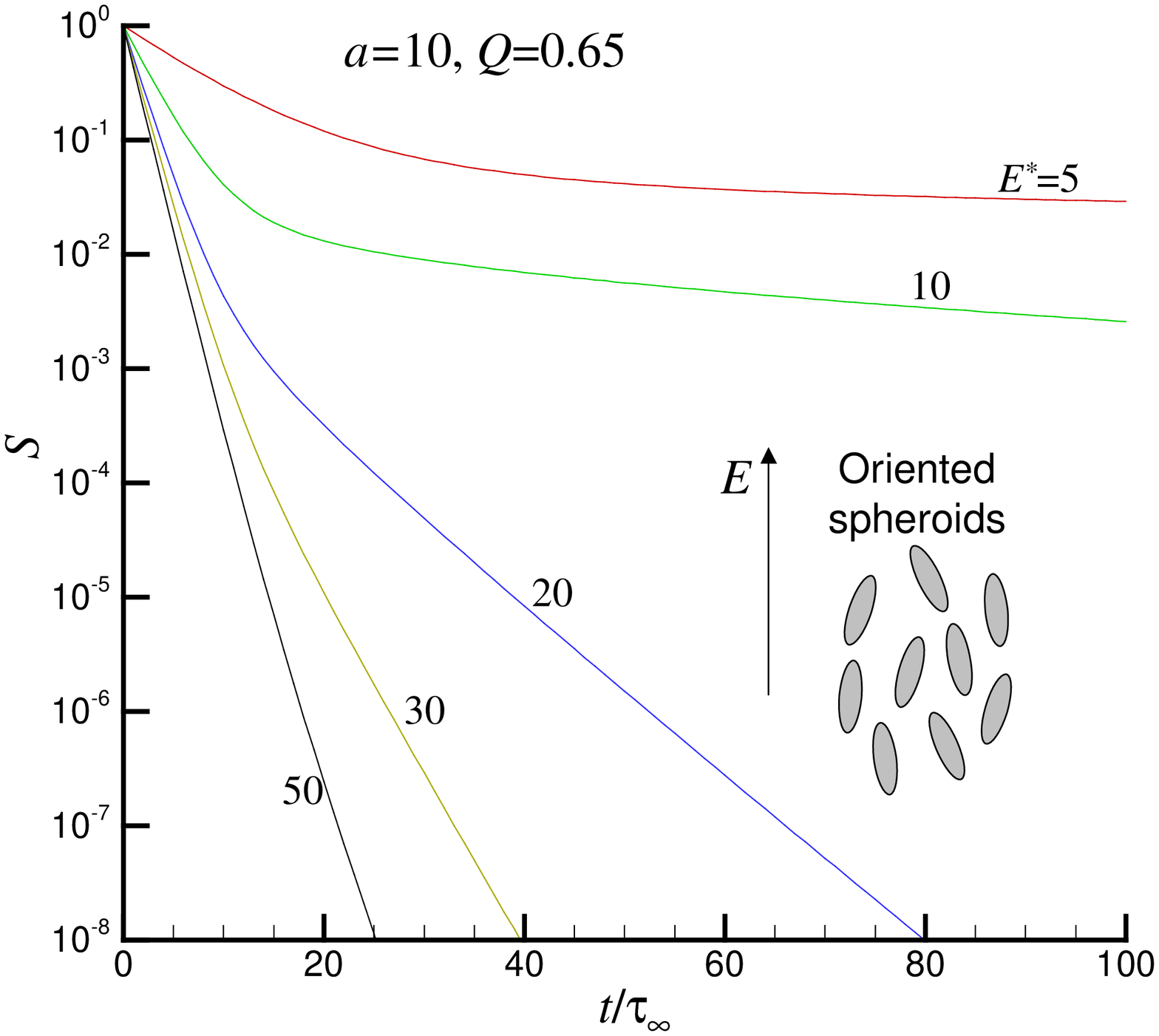}
\end{center}
\caption{Survivor curves $S ( {t/\tau _{\infty} }   )$ for
partially ordered suspensions of prolated spheroids at $a=10$,
$Q=0.65$ and different $E^{*}$ values . The dashed lines show the
slopes that correspond to orientation with $\theta $=0 and $\theta
$ =90$^{o}$.} \label{f05b}
\end{figure}

\Fref{f05a} and \fref{f05b}  show the calculated survivor curves
$S(t)$ for partially ordered suspensions of spheroidal cells. The
surviving kinetics of more disordered suspensions ($Q\to 0$) in
the limit of large time ($t/\tau _{\infty}\gg 1$) was obviously
controlled by the inactivation of cells oriented perpendicular to
the applied field, and $S(t) \approx \exp ( {-t/\tau({90^{o}})}
)$.

Increase of the order parameter $Q$ results in two different
regimes of surviving kinetics that correspond to inactivation of
cells oriented along the field (fast regime at small time
$S(t)\approx \exp(t/\tau(0^o))$ and perpendicular to the applied
field (slow regime at large time $S(t) \approx
\exp(t/\tau(90^o))$. The partial contribution of the first (fast)
regime to inactivation kinetics increases with growth of the order
parameter $Q$ (\fref{f05a}).

For partially orientationally ordered suspensions with given $Q$,
increase of the electric field strength $E^*$ causes enhancement
of inactivation kinetics and two regimes of surviving kinetics are
also observed ( \fref{f05a}). The rate of inactivation in the
regime of slow inactivation is not constant in the limit of large
time ($t/\tau_\infty\gg\ 1$) and increases with $E^*$ increase.

The orientational ordering can be induced during the PEF
treatment. Because of the quadratic dependence (see equations
\eref{e16},\eref{e17}) on the electric field strength $E$, the
orientational ordering in high pulsed electric fields may be
noticeable. The possibility of ordering for ellipsoidal or
cylindrically shaped microorganisms under the effect of external
electric fields is discussed in \cite{Asencor1993}. A rod-shaped
tobacco mosaic virus (TMV, about 0.018 $\mu $m in diameter and 0.3
$\mu $m in length \cite{Jeng1989}) demonstrates a strong
orientation, near to complete saturation of the optical
birefringence, in the electric field as high as $ \approx $4
kV/cm\cite{Konski1959}. Electrooptical studies of rod-shaped
\textit{E. coli} suspensions \cite{Khlebtsov1999, Trusov2002}
shows the existance of strong orientational ordering at electric
fields of $E<1$ kV/cm.

The orientation electric field induced effects can be roughly
estimated using \eref{e17},\eref{e18}. Taking the experimental
value of the electrical polarizability anisotropy of \textit{E.
coli} cells $\beta =4^.10^{-27}$ Fm$^2$\cite{Khlebtsov1999}, we
obtain from \eref{e17} that $U^*=\beta E^2/8\pi kT \approx
5^{.}10^3 $ at $E=1$ kV/cm and $T=298$ K. The corresponding order
parameter following from \eref{e18} is $Q\approx 1$, e.i., degree
of ordering is high.

\begin{figure}[h]
\begin{center}
\includegraphics[width=12.0cm,clip=true]{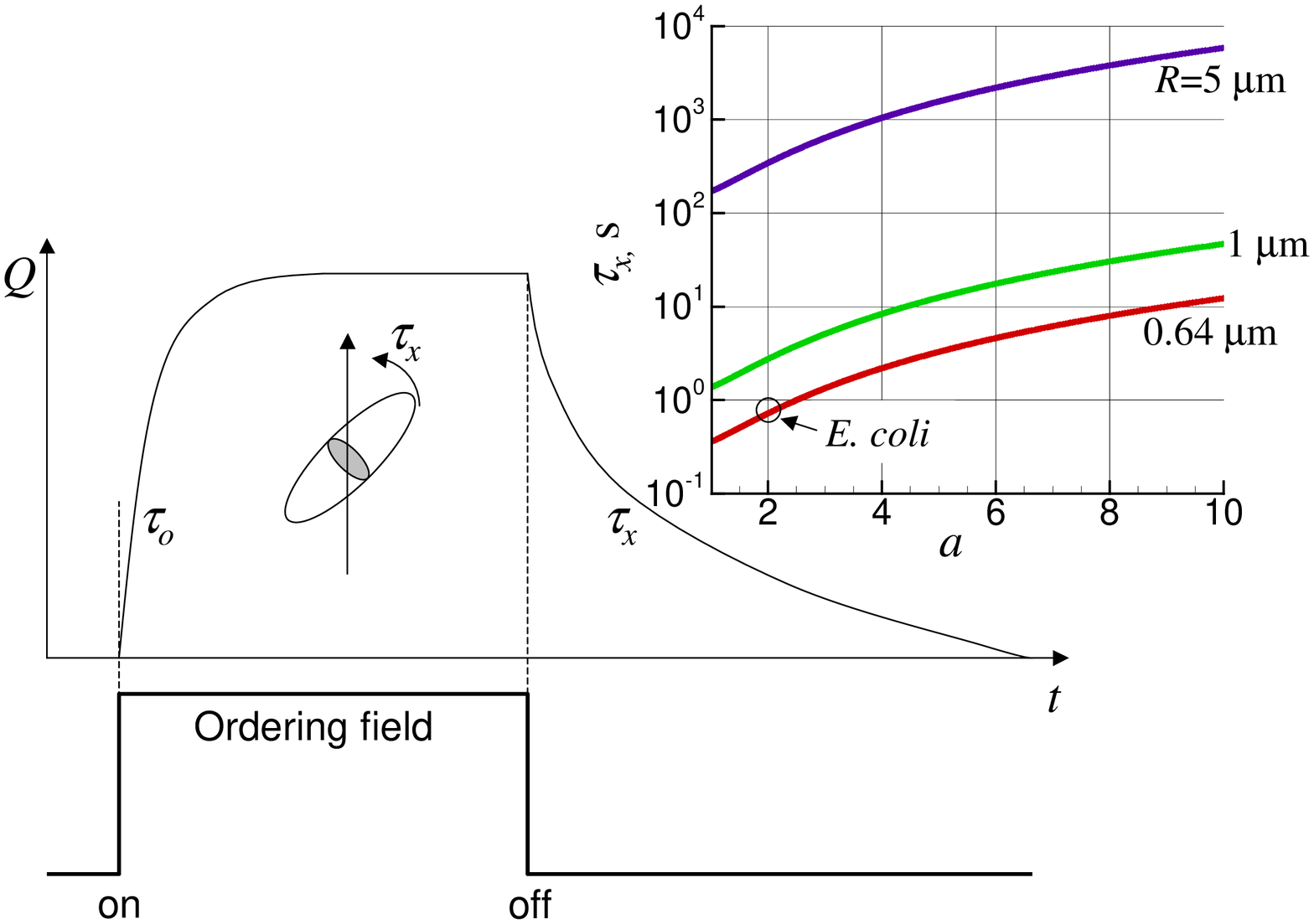}
\end{center}
\caption{Changes of order parameter $Q$ versus time $t$ in the
external electric field. Here, $\tau _{x}$ is the diffusion time
for spheroid rotation about $x$ axis, $\tau _{o}$ is the same
diffusion time in the presence of the external field. Insert shows
$\tau _{x}$ versus aspect ratio a for prolate spheroids, estimated
from \eref{e19a} and \eref{e20a}, $T=298$ K, $\eta =8.91^.10^{-4}$
Pa$^{.}$s (water viscosity \cite{Atkins1995}), $R$ is the radius
of a sphere with the same volume as spheroid.} \label{f06}
\end{figure}

But the degree of orientation can depend also on the pulse
duration. As it is shown schematically in  \fref{f06}, the order
parameter $Q$ increases with time constant $\tau_o$ after the
external electric field is switched on. The order parameter $Q$
decreases to zero with another time constant $\tau_x$ when the
external field is switched off. The relaxation time $\tau_x$ is
determined by the Brownian rotation diffusion of the spheroid
rotation about $x$ axis in absence of electric field.

According to \cite{Perrin1934}, the rotational diffusion times of
a spheroid with respect to the symmetry axes $x$ (or $y$), and $z$
are:
\begin{equation}
\tau _{x,y} = \frac{{2 ( {a^{4} - 1}  )}}{{3a ( { ( {2a^{2} - 1}
)F - a}  )}}\tau _{R} , \label{e19a}
\end{equation}
\begin{equation}
\tau _{z} = \frac{{2 ( {a^{2} - 1}  )}}{{3a ( {a - F}
 )}}\tau _{R} , \label{e19b}
\end{equation}
where $\tau _{R}$ =$\pi \eta R^{3}/kT$ is the rotational diffusion
time for a sphere of radius $R$ with the same volume as a spheroid
and
\begin{equation}
F = F_{p} = \ln ( {a + \sqrt {a^{2} - 1}}   )/\sqrt {a^{2} - 1} ,
\label{e20a}
\end{equation}
\begin{equation}
F = F_{o} = \arctan ( {\sqrt {a^{ - 2} - 1}}   )/\sqrt {1 - a^{2}}
, \label{e20b}
\end{equation}
for a prolate and oblate spheroid, respectively.

Insert in \fref{f06} shows a rotation diffusion time $\tau _{x}$
(along the short axis $x$) of the prolate spheroid  versus an
aspect ratio \textit{a} as calculated from \eref{e19a},
\eref{e20a}. For example, the rotational diffusion time is of the
order of $\tau_x \approx $1s for a \textit{E. coli}  cell with
equivolume radius $R \approx 0.64 \mu $m and aspect ratio $a
\approx 2$ \cite{Khlebtsov1999}, but it can increase substantially
with increase of the equivolume radius $R$ or aspect ratio $a$.

It is more difficulty to calculate the time $\tau _{o}$
characterizing the process of ordering in the external electric
field . The estimations shows \cite{Levy2003}that $\tau _{o}
\approx \tau _{x} $ in a low field, when, $Q<0.1$. At a very high
electric field, when $ Q \to 1.0$, $\tau_o \approx
\tau_x/(2U^*/15)$, where $U^*=\beta E^2/(8\pi kT)$ (equation
 \eref{e17}).

For rather small \textit{ E. coli}  cells from the above
estimations, we obtain $\tau _o \approx
\tau_x/(2U^*/15)\approx10^{-3}$s at $E_o =1$ kV/cm and $\tau_o
\approx 10^{-5} $s at $E =10$ kV/cm ($T=298$ K). So, small cells
like those of \textit{E. coli} may be effectively oriented during
the pulse duration at PEF treatment with field strength $E=10$
kV/cm and pulse duration $t_i=10^{-5}$s. This estimation is in
accordance with experimental observations \cite{Eynard1998}
showing that the 24 ms pulse causes the observable orientation of
the \textit{E. coli} cells parallel to the field direction only at
fields $E$ exceeding 1.25 kV/cm.

However, the PEF-induced orientation effects can be supressed for
larger cells or in presence of cells aggregation. The factor of
bacterial aggregation is essential, because particles of the large
colloidal aggregates are bounded and can not be freely reoriented
during the PEF treatment. The aggregation and colony formation is
a typical phenomenon in biocolloidal suspensions of bacterial
particles, and some kind of bacteria (e.g. \textit{Bacillus
subtilis}) exhibit various aggregation patterns
\cite{Matsushita1997,Eiha2002}. Moreover, the PEF-induced
cell-cell aggregation are also inportant \cite{Antov2005}.

For larger cells, for example, with equivolume radius $R$ of
$\approx 5.0\mu $m the rotation diffusion time is $\tau_x\approx
10^2-10^3$ s (insert in \fref{f06}). In this case $\tau_o \approx
\tau_ x/(2U^*/15)\approx 10^{-2}-10^{-3}$ s at $E=10$ kV/cm and
$t_i\ll \tau_o $. So, PEF-induced orientational effects can be
rather small and not contribute into inactivation kinetics for
large cells or cell aggregates.

\section{Conclusion and outlook}

This study has demonstrated the possibility of non-exponential
survaving kinetics of microbial inactivation under the PEF
treatment, which is believed to be related to the orientational
disorder existing in a suspension of spheroidal microbial cells.
Another factor can be related with sizes distribution of microbial
cells \cite{Lebovka2004}. Deviations from the ideal first order
kinetics law (\eref{e13}) are more pronounced in completely
disordered suspensions of prolate spheroids at small electric
field strength or at large aspect ratio $a$. Efficiency of
inactivation enhances with order parameter and field strength
increase in partially oriented suspensions. In general case the
inactivation kinetics can be influenced by the concentration of
cells, their aggregation and arrangement \cite{Canatella2004,
Pucihar2007, Pavlin2007}. The relevance and importance of such
effects for explanation of the survival curves observed in
PEF-inactivation experiments should be studied in future. The
accurate description of inactivation kinetics requires accounting
for the dynamics of bacterial cell reorientation in a high
electric field during its inactivation. It seems to be important
also to find correlations between factors that controls
aggregations of bacterial cells, PEF protocols and parameters of
bacterial inactivation kinetics.

\section{Acknowledgments}
%\begin{acknowledgments}
The authors would like to thank the "Pole Regional Genie des
Procedes" (Picardie, France) for providing the financial support.

%\end{acknowledgments}

\section*{Figure legends}

%---------------------------------------------------------------------------
%---------------------------------------------------------------------------
\newpage
%---------------------------------------------------------------------------
%---------------------------------------------------------------------------
\newpage
%---------------------------------------------------------------------------
%---------------------------------------------------------------------------

\newpage
%---------------------------------------------------------------------------
%---------------------------------------------------------------------------
\newpage
%---------------------------------------------------------------------------
%---------------------------------------------------------------------------
\newpage
%---------------------------------------------------------------------------
%---------------------------------------------------------------------------
\newpage
%---------------------------------------------------------------------------
%---------------------------------------------------------------------------
\newpage
%---------------------------------------------------------------------------
%---------------------------------------------------------------------------
\newpage
%---------------------------------------------------------------------------
%---------------------------------------------------------------------------
\newpage
%---------------------------------------------------------------------------
%---------------------------------------------------------------------------

\end{document}